\documentclass[12pt]{article}

\usepackage[T2A]{fontenc}
\usepackage[utf8]{inputenc}
\usepackage[english]{babel}
\usepackage{indentfirst}

\usepackage{bm}
\usepackage{mathrsfs}
\usepackage{amsmath}
\usepackage{amssymb}
\usepackage{mathtools}
\usepackage{enumitem}
\usepackage{cite}
\usepackage{hyperref}

\DeclareMathOperator*{\Iunit}{i}

\renewcommand{\imath}{\Iunit}

\makeatletter
\let\orig@phi\phi
\let\orig@varphi\varphi
\let\orig@epsilon\epsilon
\let\orig@varepsilon\varepsilon
\let\orig@kappa\kappa
\let\orig@varkappa\varkappa
\renewcommand{\phi}{\orig@varphi}
\renewcommand{\varphi}{\orig@phi}
\renewcommand{\epsilon}{\orig@varepsilon}
\renewcommand{\varepsilon}{\orig@epsilon}
\renewcommand{\kappa}{\orig@varkappa}
\renewcommand{\varkappa}{\orig@kappa}
\makeatother

\allowdisplaybreaks[4]

\begin{document}

\begin{center}
  \bfseries\large
  The neutrino propagator in matter and its spin properties
\end{center}

\begin{center}{D.\,M.\,Voronin\footnote{e-mail: dmitry.m.voronin@gmail.com},
	A.\,E.\,Kaloshin\footnote{e-mail: alexander.e.kaloshin@gmail.com}}
\end{center}
\begin{center}
	\textit{Irkutsk state University, 664003, Irkutsk, Russia}
\end{center}	

\vspace*{2em}

\begin{abstract}
 A spectral representation for the neutrino propagator in a moving matter with constant density is constructed. It is found that in a matter there exists a 4-dimensional axis of complete polarization, all poles of the propagator are classified according to the values of the spin projection onto this axis.
\end{abstract}


	\section{Introduction} 
	Neutrino physics at present is becoming a tool of research in various fields of physics: from geophysics to physics beyond the Standard Model. If to say about propagation of neutrino through a matter, the most important effect is related with resonance amplification of oscillations \cite{q16,m5656} and problem of deficit of solar neutrino, see reviews \cite{q8,Der14}. Either quantum-mechanical or field-theoretic approaches are being used for description of the propagation of neutrino in matter, see, for example \cite{m56,647}.
	
	In this paper we construct a spectral representation of propagator in moving matter, which is based on the eigenvalue problem for inverse propagator. It turns that it is possibly to write a simple analytical solution in general case, without any restrictions for the form of interaction. The spectral representation gives a maximally simple algebraic approach for description of neutrino propagation in moving matter in the quantum field theory.

	
	
	\section{The propagator in matter and spin projectors} 
	Propagator of fermion in a matter is central object in consideration of problem in quantum field theory approach. It contains two four-dimensional vectors --- momentum $p$ and velocity of matter $u$, and one can write only 8 $\gamma$-matrix structures in the decomposition with account of parity violation. The most general expression for the inverse propagator looks as:
	\begin{equation}\label{formula2}
	S=G^{-1} =s_{1}I+s_{2}\hat{p}+s_{3}\hat{u}+s_{4}\sigma^{\mu\nu}p_{\mu}u_{\nu}+s_{5}i\varepsilon^{\mu\nu\lambda\rho}\sigma^{\mu\nu}u_{\lambda}p_{\rho}+s_{6}\gamma^{5}+s_{7}\hat{p}\gamma^{5}+s_{8}\hat{u}\gamma^{5},
	\end{equation}
	where $s_{i}$ are the scalar functions depending on invariants.
	
	Below we solve the eigenvalue problem for the inverse propagator of the general form. It is convenient to use instead of $\gamma$-matrix basis another one with simple multiplicative properties.
	Following  \cite{q11}, let us introduce the vector $z^{\mu}$ which is a linear combination of the vectors $p$, $u$ and has the properties of the polarization vector of the fermion:
	\begin{equation}\label{123sa}  
	z^{\mu}p_{\mu}=0,  \ \ \ z^{2}=-1. 
	\end{equation} 
	It has form:
	\begin{equation}\label{bvbfe}
	z^{\mu}=b(p^{\mu}(up)-u^{\mu}p^{2}), ~~b=1/\sqrt{p^{2}[(up)^{2}-p^{2}]}.
	\end{equation}
	
	Using this vector, we construct generalized spin projectors
	\footnote{We call them generalized because of the additional factor $\hat{n}$. But in fact \eqref{Sigma} is the most common form of spin projectors with parity violation and in presence of mixing --- see details in \cite{q98}.}:
	\begin{equation}\label{Sigma}
	\varSigma^{\pm}=\dfrac{1}{2}(1\pm\gamma^{5}\hat{z}\hat{n}),~~~ \varSigma^{\pm}\varSigma^{\pm}=\varSigma^{\pm},~~~\varSigma^{\pm}\varSigma^{\mp}=0,
	\end{equation}
	where $n^{\mu}=\dfrac{p^{\mu}}{W},\ \  W=\sqrt{p^{2}}$. 
	
	It is easy to check that $\varSigma^{\pm}$ commute with inverse propagator (\ref{formula2}).
	
	If to multiply $S(p,u)$ \eqref{formula2} by unit matrix:
	\begin{equation}
	S=(\Sigma^{+}(z)+\Sigma^{-}(z))S \equiv S^+ + S^- ,
	\end{equation}
	we will have two terms orthogonal to each other.
	
	Another property of $\varSigma^{\pm}$ is that "under observation" of the spin projector (i.e. in terms of $S^+, S^-$) $\gamma$-matrix structures are greatly simplified. Namely, $\gamma$-matrices with 4-velocity $u^{\mu}$ are reduced to set of $\gamma$-matrices: $I, \hat{p}, \gamma^{5}, \hat{p}\gamma^{5}$.
	
	For example, term $\hat{u}$ in \eqref{formula2} can be rewritten as a linear combination of $\hat{p}$ and $\hat{z}$ and then simplified in presence of the projector:
	\begin{equation}
	\dfrac{1+\gamma^{5}\hat{z}\hat{n}}{2}\hat{u}=\dfrac{1+\gamma^{5}\hat{z}\hat{n}}{2}(a_1\hat{p}+a_2\hat{z})=\Sigma^{+}(z)(a_1\hat{p}-\dfrac{a_2}{W}\hat{p}\gamma^{5}).
	\end{equation}
	
	After this simplification it is convenient to use the off-shell projectors:
	\begin{equation}
	\Lambda^{\pm}=\dfrac{1}{2}(1\pm\hat{n}),~~~n^{\mu}=\dfrac{p^{\mu}}{W}
	\end{equation}
	orthogonal to each other.
	
	Having a set of projectors $\Lambda^{\pm}$ and $\varSigma^{\pm}$ one can construct a basis which we will be used below:
	\begin{eqnarray}\label{1}
	R_{1}=\varSigma^{-}\varLambda^{+},~~~~~~R_{5}=\varSigma^{+}\varLambda^{+},~~~\nonumber\\
	R_{2}=\varSigma^{-}\varLambda^{-},~~~~~~R_{6}=\varSigma^{+}\varLambda^{-},~~~\nonumber\\
	R_{3}=\varSigma^{-}\varLambda^{+}\gamma^{5},~~~R_{7}=\varSigma^{+}\varLambda^{+}\gamma^{5},\nonumber\\
	R_{4}=\varSigma^{-}\varLambda^{-}\gamma^{5},~~~R_{8}=\varSigma^{+}\varLambda^{-}\gamma^{5}.
	\end{eqnarray} 
	Multiplicative properties of a basis \eqref{1} are given in tab.\eqref{lklhjlksd}, where the element from column is multiplied from left on the element of row.
	
	The inverse propagator \eqref{formula2} can be decomposed in this basis
	\begin{equation}\label{decomp}
	S(p,u)= \sum_{i=1}^{4} R_i S_i (p^2,pu) + \sum_{i=5}^{8} R_i S_i (p^2,pu),
	\end{equation}
	and these two sums are orthogonal to each other.
	
	\begin{table}[h]
		\caption{Multiplicative properties of elements of the basis \eqref{1}}\label{lklhjlksd}
		\begin{center}
			\begin{tabular}{|c|c|c|c|c||c|c|c|c|c|} \hline      
				&$R_{1}$&$R_{2}$&$R_{3}$&$R_{4}$&$R_{5}$&$R_{6}$&$R_{7}$&$R_{8}$ \\ \hline
				$R_{1}$&$R_{1}$&0&$R_{3}$&0&0&0&0&0\\ 
				$R_{2}$&0&$R_{2}$&0&$R_{4}$&0&0&0&0\\   
				$R_{3}$&0&$R_{3}$&0&$R_{1}$&0&0&0&0\\ 
				$R_{4}$&$R_{4}$&0&$R_{2}$&0&0&0&0&0\\ \hline\hline
				$R_{5}$&0&0&0&0&$R_{5}$&0&$R_{7}$&0\\ 
				$R_{6}$&0&0&0&0&0&$R_{6}$&0&$R_{8}$\\ 
				$R_{7}$&0&0&0&0&0&$R_{7}$&0&$R_{5}$\\ 
				$R_{8}$&0&0&0&0&$R_{8}$&0&$R_{6}$&0\\ \hline
			\end{tabular}
		\end{center}
	\end{table}
	
	
	The table shows that in this case the eigenvalue problem for inverse propagator is separated into two problems for $R_{1}..R_{4}$ and $R_{5}..R_{8}$. Each of these problems has two eigenvalues.
	
	\section{Spectral representation of the propagator} 
	
	Let us consider eigenvalue problem for some linear hermitian operator
	\begin{equation}
	\hat{A}|\Psi_{i}\rangle=\lambda_{i}|\Psi_{i}\rangle .
	\end{equation}
	Having solved this problem, we can represent the operator as a spectral decomposition \cite{q10}
	\begin{equation}
	\hat{A}=\sum\lambda_{i}|\Psi_{i}\rangle\langle\Psi_{i}| =\sum\lambda_{i} \Pi_{i},
	\end{equation}
	which contains eigenvalues $\lambda_{i}$ and eigenpojectors $\Pi_{i} =|\Psi_{i}\rangle\langle\Psi_{i}|$. Orthonormality of eigenvectors leads to the orthogonality property of projectors
	\begin{equation}
	\Pi_{i}\Pi_{k}=\delta_{ik}\Pi_{k} .
	\end{equation} 
	If operator is non-hermitian, than to construct a spectral decomposition we need to solve two eigenvalue problems: left and right.

	We want to construct a spectral representation for the inverse propagator in the most general form \eqref{formula2} or \eqref{decomp}, so we need to solve the eigenvalue problem
	\begin{equation}\label{eq:5}
	S\Pi_i = \lambda_i \Pi_i .
	\end{equation}
	Note that we prefer to solve the problem in a matrix form, i.e. we are looking for enprojectors $\Pi_i$ from the begining. This can be done with use of  matrix basis, and this allows to avoid  cumbersome intermediate formulas. As for non-hermitiancy of propagator: it is enough to solve the left eigenvalue problem and to require an orthogonality of the obtained projectors -- see \cite{q12}.
	
	After solving this problem, we have a spectral representation for the inverse propagator in a moving matter:
	\begin{equation}\label{RS}
	S(p,u)=\sum_{i=1}^{4} \lambda_{i}\Pi_{i}. 
	\end{equation}
	
	The propagator is obtained by reversing of \eqref{RS} and if the eigenprojectors $\Pi_{i}$ form a complete orthogonal system the answer is:
	\begin{equation}\label{spec_G}
	G(p,u)=\sum_{i=1}^{4} \frac{1}{\lambda_{i}}\Pi_{i}. 
	\end{equation}
	
	The use of the basis \eqref{1} essentially simplifies the solving of eigenvalue problem. The matrix solutions  also can  be written as a decomposition in this basis, so the orthogonality of the spin projectors (see Table) leads to a simpler
	problems involving $R_{1}..R_{4}$ and $R_{5}..R_{8}$
	in \eqref{decomp}. It was noted above that "under observation" of spin projector the gamma-matrices in \eqref{formula2} are turned into a set of $I, \hat{p}, \gamma^{5}, \hat{p}\gamma^{5}$. So, for example, the eigenvalue problem for the first four terms
	\begin{equation}\label{RS1}
	\left( \sum_{k=1}^{4}R_{k}  S_{k} \right)\cdot \left( \sum_{i=1}^{4}R_{i}  A_{i} \right)  = \lambda \left( \sum_{i=1}^{4}R_{i}  A_{i} \right)   
	\end{equation}  
	algebraically repeats a similar problem for dressed vacuum propagator with parity violation \cite{q12}. Presence of matter is revealed in appearance of spin projectors and in scalar coefficients of decomposition. Besides, in a matter there is a doubling of the number of eigenvalues, we have two quadratic equations for them.

	Repeating the algebraic actions \cite{q12} one can write the answer for the eigenvalue problem in the general case. Eigenprojectors and eigenvalues for inverse propagator look as:  
	\begin{equation*}
	\Pi_{1}=\dfrac{1}{\lambda_{2}-\lambda_{1}}\Big((S_{2}-\lambda_{1})R_{1}+(S_{1}-
	\lambda_{1})R_{2}-S_{3}R_{3}-S_{4}R_{4}\Big),
	\end{equation*}
	\begin{equation*}
	\Pi_{2}=\dfrac{1}{\lambda_{1}-\lambda_{2}}\Big((S_{2}-\lambda_{2})R_{1}+(S_{1}-
	\lambda_{2})R_{2}-S_{3}R_{3}-S_{4}R_{4}\Big),
	\end{equation*}
	\begin{equation*}
	\Pi_{3}=\dfrac{1}{\lambda_{4}-\lambda_{3}}\Big((S_{6}-\lambda_{3})R_{5}+(S_{5}-
	\lambda_{3})R_{6}-S_{7}R_{7}-S_{8}R_{8}\Big),
	\end{equation*} 
	\begin{equation}\label{eig_pro}
	\Pi_{4}=\dfrac{1}{\lambda_{3}-\lambda_{4}}\Big((S_{6}-\lambda_{4})R_{5}+(S_{5}-
	\lambda_{4})R_{6}-S_{7}R_{7}-S_{8}R_{8}\Big),
	\end{equation}
	\begin{equation*}
	\lambda_{1,2}=\dfrac{S_{1}+S_{2}}{2}\pm\sqrt{\Big(\frac{S_{1}-S_{2}}{2}\Big)^{2} + S_{3}S_{4}}~,
	\end{equation*}
	\begin{equation}
	\label{eig_val}
	\lambda_{3,4}=\dfrac{S_{5}+S_{6}}{2}\pm\sqrt{\Big(\frac{S_{5}-S_{6}}{2}\Big)^{2} + S_{7}S_{8}}~.
	\end{equation} 
	Here $S_{i}$ are coefficients of decomposition of inverse propagator in the basis (\ref{decomp}).
	
	The obtained projectors have the following properties:
	\begin{enumerate}
		\item$S\Pi_{k}=\lambda_{k}\Pi_{k}$, 
		\item$\Pi_{i}\Pi_{j}=\delta_{ij}\Pi_{j}$,
		\item$\sum\limits_{i=1}^4\Pi_{i}=1$.
	\end{enumerate}
	
	\subsection{Propagator in the Standard Model} 
	
	The inverse propagator of fermion in the Standard Model in matter looks as:
	\begin{equation}\label{dddf}
	S=\hat{p}-m+\alpha\hat{u}(1-\gamma^{5}),
	\end{equation}
	where $u$ is 4-vector of velocity of matter, $p$ --- 4-vector of momentum of particle and $\alpha$ is a parameter depending of densities of electrons, protons and neutrons in a matter.
	
	
	Let us write down the coefficients of decomposition in two bases: $\gamma$-matrix \eqref{formula2} and $R$-basis (\ref{decomp}):
	\begin{eqnarray}\label{vvbnmkl}
	s_{1}=-m,~~~A_{1}=-m+W(1+K^{+});\nonumber\\
	s_{2}=1,~~~~~~~A_{2}=-m-W(1+K^{+});\nonumber\\
	s_{3}=\alpha,~~~~~~~A_{3}=-WK^{+};~~~~~~~~~~~~~~\nonumber\\
	s_{4}=0,~~~~~~~~A_{4}=WK^{+};~~~~~~~~~~~~~~~~\nonumber\\
	s_{5}=0,~~~~~~~A_{5}=-m-W(1+K^{-});\nonumber\\
	s_{6}=0,~~~~~~~A_{6}=-m+W(1+K^{-});\nonumber\\
	s_{7}=0,~~~~~~~~A_{7}=WK^{-};~~~~~~~~~~~~~~~~\nonumber\\
	s_{8}=-\alpha,~~~~~A_{8}=-WK^{-}.~~~~~~~~~~~~~~
	\end{eqnarray}
	Here  notations are introduced:
	$K^{\pm}=\dfrac{\alpha}{W^{2}}\Big((pu)\pm\sqrt{(up)^{2}-W^{2}}\Big)$,   $W=\sqrt{p^{2}}$.
	
	Solutions of eigenvalue problem (\ref{eq:5}) in this case are:
	\begin{equation}\label{mnn}
	\lambda_{1,2}=-m \pm W\sqrt{1+2K^{+}}
	\end{equation}
	\begin{equation}
	\lambda_{3,4}=-m \pm W\sqrt{1+2K^{-}}
	\end{equation}
	\begin{equation}
	\Pi_{1,2}=\varSigma^{-}\cdot \dfrac{1}{2} \left[1\pm \hat{n}\ \frac{1+K^{+} - \gamma^{5}K^{+}}{\sqrt{1+2K^{+}}}  \right]
	\end{equation}
	\begin{equation}\label{vddaq}
	\Pi_{3,4}=\varSigma^{+}\cdot \dfrac{1}{2} \left[1\pm \hat{n}\ \frac{1+K^{-} - \gamma^{5}K^{-}}{\sqrt{1+2K^{-}}}  \right]
	\end{equation}

	\subsection{
		Case of rest matter} We consider a special case of propagator \eqref{dddf}, namely situation, when matter is at rest ($\vec{u}=0, u_{0}=1$). In this case vector of polarization $z^\mu$ (\ref{bvbfe})  takes form 
	\begin{equation}\label{}
	z^{\mu}=\frac{1}{W} \left( |\vec{p}| ,\ \frac{\vec{p} p^0}{|\vec{p}|}  \right), 
	\end{equation}
	which corresponds to the helicity state of the fermion, but off-mass-shell, since $W\not= m$. The spin projectors in (\ref{Sigma}) are projectors onto the  the 3-momentum
	\begin{equation}\label{}
	\Sigma^{\pm}=\frac{1}{2} \left( 1\pm \vec{\Sigma} \frac{\vec{p}}{|\vec{p}|} \right), \  \   \vec{\Sigma}= \gamma^0 \vec{\gamma} \gamma^5 .
	\end{equation}
	
	In above formulas for eigenvalues and eigenprojectors  the factors $K^\pm$ are simplified in the following way:
	\begin{equation}\label{}
	K^{\pm}=\dfrac{\alpha}{W^{2}}\Big( p^0 \pm |\vec{p}|\Big).
	\end{equation}
	
	Thus, in case of the rest matter the well-known fact \cite{q13, q89} is reproduced, that neutrino with definite helicity has a definite law of dispersion.
	
	\section{Conclusion}
	So, we have constructed spectral representation of the propagator (\ref{spec_G}), (\ref{eig_pro}), (\ref{eig_val}) of neutrino in a moving matter. In this form, based on the eigenvalue problem, the propagator is represented as a sum of poles with its $\gamma$-matrix projector and each term is related to a definite dispersion law in matter. More precisely, the relation of energy and momentum arises as a result of the eigenvalues equality (??) to zero $\lambda_i = 0$ in (\ref{spec_G}). Solutions of eigenvalue problem are obtained for propagator of the most general form (\ref{formula2}).
	Let us stress a special role of generalized spin projectors (\ref{Sigma}) onto the fixed axis (\ref{bvbfe}), its appearence essentially simplifies all algebraic properties.

	In the case of a moving matter, the states with definite spin-to-axis projection (\ref{bvbfe}) have a definite dispersion law. In the particular case of a rest matter, the operators $\varSigma^{\pm}$ are projectors onto states with a certain helicity, which corresponds to the earlier known results \cite{q13,q89}. Note that for a moving matter the zero commutator $[S,\Sigma^{\pm}] = 0$ does not mean a conservation of spin projection onto this axis, since the projectors $\Sigma^{\pm}$ do not commute with the Hamiltonian.
	
	We have considered the case of a moving unpolarized matter. Account of polarization leads to a simple replacement of $u^{\mu}$ by a combination of vectors of velocity and polarization of matter \cite{zas}, the method of calculations is not changed.
	
	The constructed spectral representation of the propagator is the simplest and most convenient algebraic construction for describing the effects of mixing and neutrino oscillations in a matter. Note that the appearence of spin projectors will  essentially simplify this problem also and the algebraic construction will repeat the problem of mixing in a vacuum \cite{q98}. Possible applications of the developed approach are related with astrophysical problems, in particular, when the neutrino propagates through the envelope of the supernova, see e.g. review \cite{zas1}. The investigation of the spin properties, with account the discovered off-shell axis of complete polarization, deserves a separate consideration.
	
	We are grateful to V. A. Naumov for useful discussion.

\end{document}